\pdfoutput=1
\documentclass[my,twocolumn]{webofc}

\usepackage{graphicx}
\usepackage{xspace}
\usepackage{bm,color}

\def\lhcb{\mbox{LHCb}\xspace}

\def\order{\ensuremath{\mathcal{O}}\xspace}
\def\CP{\ensuremath{C\!P}\xspace}
\def\squark{\ensuremath{s}\xspace}

\def\bquark{\ensuremath{b}\xspace}
\def\PK{\ensuremath{K}\xspace}
\def\PB{\ensuremath{B}\xspace}
\def\PD{\ensuremath{D}\xspace}
\def\Kbar{\kern 0.2em\overline{\kern -0.2em \PK}{}\xspace}
\def\Kz{\ensuremath{K^0}\xspace}
\def\Kzb{\ensuremath{\Kbar^0}\xspace}
\def\Bbar{\kern 0.18em\overline{\kern -0.18em \PB}{}\xspace}
\def\Bz{\ensuremath{B^0}\xspace}
\def\Bzb{\ensuremath{\Bbar^0}\xspace}
\def\Bs{\ensuremath{B^0_\squark}\xspace}
\def\Bsb{\ensuremath{\Bbar^0_\squark}\xspace}
\def\Dbar{\kern 0.2em\overline{\kern -0.2em \PD}{}\xspace}
\def\D{\ensuremath{\PD}\xspace}
\def\Dz{\ensuremath{\D^0}\xspace}
\def\Dzb{\ensuremath{\Dbar^0}\xspace}
\def\Dstarp{\ensuremath{\D^{*+}}\xspace}
\def\invfb{\ensuremath{\mbox{\,fb}^{-1}}\xspace}
\newcommand{\tev}{\ensuremath{\mathrm{\,Te\kern -0.1em V}}\xspace}
\newcommand{\gev}{\ensuremath{\mathrm{\,Ge\kern -0.1em V}}\xspace}
\newcommand{\mev}{\ensuremath{\mathrm{\,Me\kern -0.1em V}}\xspace}

\newcommand{\massmev}{\mbox{\mev/$c^2$}}
\newcommand{\pgev}{\mbox{\gev/$c$}}

\newcommand{\pis}{\ensuremath{\pi_{\rm s}}\xspace}
\newcommand{\M}{\ensuremath{M(\Dz\pis^+)}\xspace}

\graphicspath{{./figs/}}

\usepackage[varg]{txfonts}   
\woctitle{-}
\begin{document}
\title{Charm mixing at LHCb}
\author{Angelo Di Canto\inst{1}}
\institute{Physikalisches Institut, Ruprecht-Karls-Universit\"{a}t Heidelberg, Germany}
\abstract{We report a measurement of the time-dependent ratio of $\Dz\to K^+\pi^-$ to $\Dz\to K^-\pi^+$ decay rates in \Dstarp-tagged events using $1.0\invfb$ of integrated luminosity recorded by the LHCb experiment. We measure the mixing parameters \mbox{$x'^2=(-0.9\pm1.3)\times10^{-4}$}, \mbox{$y'=(7.2\pm2.4)\times10^{-3}$} and the ratio of doubly-Cabibbo-suppressed to Cabibbo-favored decay rates \mbox{$R_D=(3.52\pm0.15)\times10^{-3}$}. The result excludes the no-mixing hypothesis with a probability corresponding to $9.1$ standard deviations and represents the first observation of charm mixing from a single measurement.}
\maketitle
\section{Introduction}
Quantum-mechanical mixing between neutral meson particle and antiparticle flavour eigenstates provides important information about electroweak interactions and the Cabibbo-Kobayashi-Maskawa matrix, as well as the virtual particles that are exchanged in the mixing process itself. For this reason, the mixing of neutral mesons is generally considered a powerful probe to discover physics beyond the standard model. Meson-antimeson mixing has been observed in the $\Kz-\Kzb$ \cite{Lande:1956pf}, $\Bz-\Bzb$ \cite{Albrecht:1987dr} and $\Bs-\Bsb$ \cite{Abulencia:2006ze} systems, all with rates in agreement with standard model expectations. Evidence of mixing in the charm system has been reported by three experiments using different \Dz decay channels~\cite{Aubert:2007wf,Staric:2007dt,Aaltonen:2007ac,Aubert:2007en,Aubert:2008zh,Aubert:2009ai}. Only the combination of these measurements provides confirmation of $\Dz-\Dzb$ mixing with more than $5\sigma$ significance \cite{hfag}. While it is accepted that charm mixing occurs, a clear observation of the phenomenon from a single measurement is needed to establish it conclusively. 

Thanks to the large charm production cross-section available in $pp$ collisions at $\sqrt{s}=7\,\tev$ and to its flexible trigger on hadronic final states, the \lhcb experiment ~\cite{Alves:2008zz}  collected the world's largest sample of fully reconstructed hadronic charm decays during the 2011 LHC run. In the following we present the first search for $\Dz-\Dzb$ mixing using this data sample, which corresponds to $1.0\,\text{fb}^{-1}$ of integrated luminosity. More details on the analysis can be found in Ref.~\cite{preprint}.

\section{Measurement of charm mixing in the $D^0\to K^+\pi^-$ channel}
In this analysis a search for $\Dz-\Dzb$ mixing is performed by measuring the time-dependent ratio of $\Dz\to K^+\pi^-$ to $\Dz\to K^-\pi^+$ decay rates.\footnote{Charge-conjugated modes are implied unless otherwise stated.} The \Dz flavour at production time is determined using the charge of the soft (low-momentum) pion, $\pis^+$, in the strong $\Dstarp\to\Dz\pis^+$ decay. The \mbox{$\Dstarp\to\Dz(\to K^-\pi^+)\pis^+$} process is referred to as right-sign (RS), whereas the \mbox{$\Dstarp\to\Dz(\to K^+\pi^-)\pis^+$} is designated as wrong-sign (WS). The RS process is dominated by a Cabibbo-favored (CF) decay amplitude, whereas the WS amplitude includes contributions from both the doubly-Cabibbo-suppressed (DCS) $\Dz\to K^+\pi^-$ decay, as well as $\Dz-\Dzb$ mixing followed by the favored $\Dzb\to K^+\pi^-$ decay. In the limit of small mixing ($|x|,|y|\ll1$), and assuming negligible \CP violation, the time-dependent ratio, $R(t)$, of WS to RS decay rates is approximated by
\begin{equation}\label{eq:true-ratio}
R(t) \approx R_D+\sqrt{R_D}\ y'\ \frac{t}{\tau}+\frac{x'^2+y'^2}{4}\left(\frac{t}{\tau}\right)^2,
\end{equation}
where $t/\tau$ is the decay time expressed in units of the average \Dz lifetime $\tau$, $R_D$ is the ratio of DCS to CF decay rates, $x'$ and $y'$ are the mixing parameters rotated by the strong phase difference between the DCS and CF amplitudes. In case of no mixing, $x'=y'=0$, the WS/RS ratio would be independent of decay time.

\subsection{Outline of the analysis}
The candidate reconstruction exploits the capabilities of the silicon vertex locator (VELO), to identify the displaced two-track vertices of the $D^0$ decay products with decay-time resolution $\Delta t\approx 0.1\tau$; of the tracking system, which measures charged particles with momentum resolution $\Delta p/p$ that varies from $0.4\%$ at $5\,\pgev$ to $0.6\%$ at $100\,\pgev$, corresponding to a typical mass resolution of approximately $8\,\massmev$ for a two-body charm-meson decay; and of the ring imaging Cherenkov detectors, which are used to distinguish between pions and kaons and to suppress the contamination from misidentified two-body charm decays in the sample. 

\begin{figure*}[t]
\centering
\includegraphics[width=0.5\textwidth]{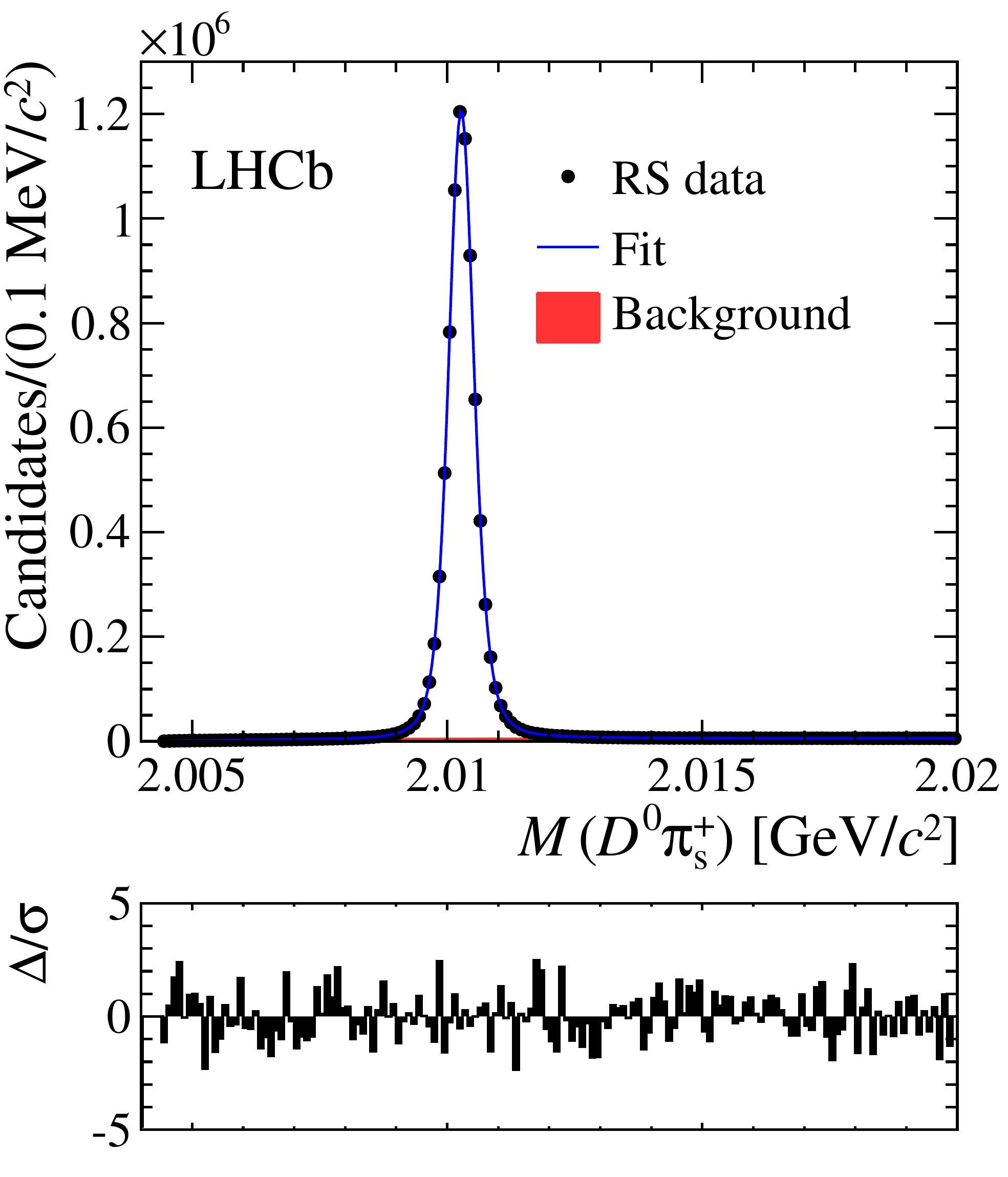}\hfil
\includegraphics[width=0.5\textwidth]{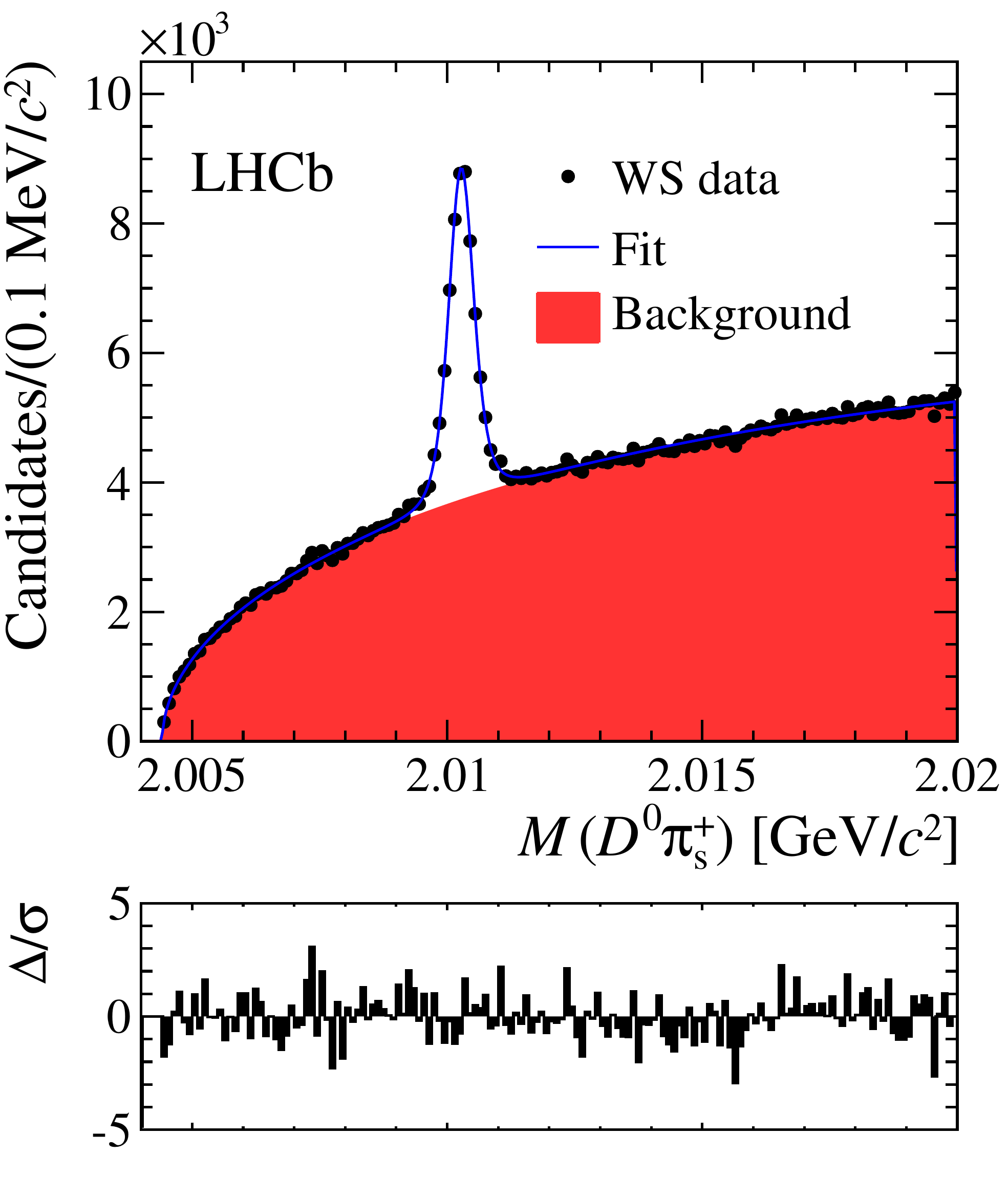}\\
\caption{Time-integrated $\Dz\pis^+$ mass distributions for the selected RS $\Dz\to K^-\pi^+$ (left) and WS $\Dz\to K^+\pi^-$ (right) candidates with fit projections overlaid. The bottom plots show the normalized residuals between the data points and the fits.}\label{fig:yields}
\end{figure*}

We reconstruct approximately $3.6\times10^4$ WS and $8.4\times10^6$ RS decays as shown in figure~\ref{fig:yields}, where the \M distribution for the selected RS and WS candidates is fitted to separate the \Dstarp signal component, with a mass resolution of about $0.3\,\massmev$, from the background component, which is dominated by associations of real \Dz decays and random pions. Similar fits are used to determine the signal yields for the RS and WS samples in thirteen \Dz decay time bins, chosen to have a similar number of candidates in each bin. The shape parameters and the yields of the two components, signal and random pion background, are left free to vary in the different decay time bins. We further assume that the \M signal shape for RS and WS decay are the same and therefore first perform a fit to the more abundant and cleaner RS sample to determine the signal shape and yield, and then use those shape parameters with fixed values when fitting for the WS signal yield. The signal yields from the thirteen bins are used to calculate the WS/RS ratios and the mixing parameters are determined in a binned $\chi^2$ fit to the observed decay-time dependence.

\subsection{Systematic uncertainties}
Since WS and RS events are expected to have the same decay-time acceptance and \M distributions, most systematic uncertainties affecting the determination of the signal yields as a function of decay time cancel in the ratio between WS and RS events. Residual biases from non-cancelling instrumental and production effects, such as asymmetries in detection efficiencies or in production, are found to modify the WS/RS ratio only by a relative fraction of $\order(10^{-4})$ and are neglected. Uncertainties in the distance between VELO sensors can lead to a bias of the decay-time scale. The effect has been estimated to be less than 0.1\% of the measured time and translates into relative systematic biases of $0.1\%$ and $0.2\%$ on $y'$ and $x'^2$, respectively. At the current level of statistical precision, such small effects are negligible.

The main sources of systematic uncertainty are those which could alter the observed decay-time dependence of the WS/RS ratio. 
Two such sources have been identified: $(1)$ \D mesons from \bquark-hadron decays, and $(2)$ peaking backgrounds from charm decays reconstructed with the wrong particle identification assignments. These effects, discussed below, are expected to depend on the true value of the mixing parameters and are accounted for in the time-dependent fit.

\begin{figure*}[t]
\centering
\includegraphics[width=0.48\textwidth]{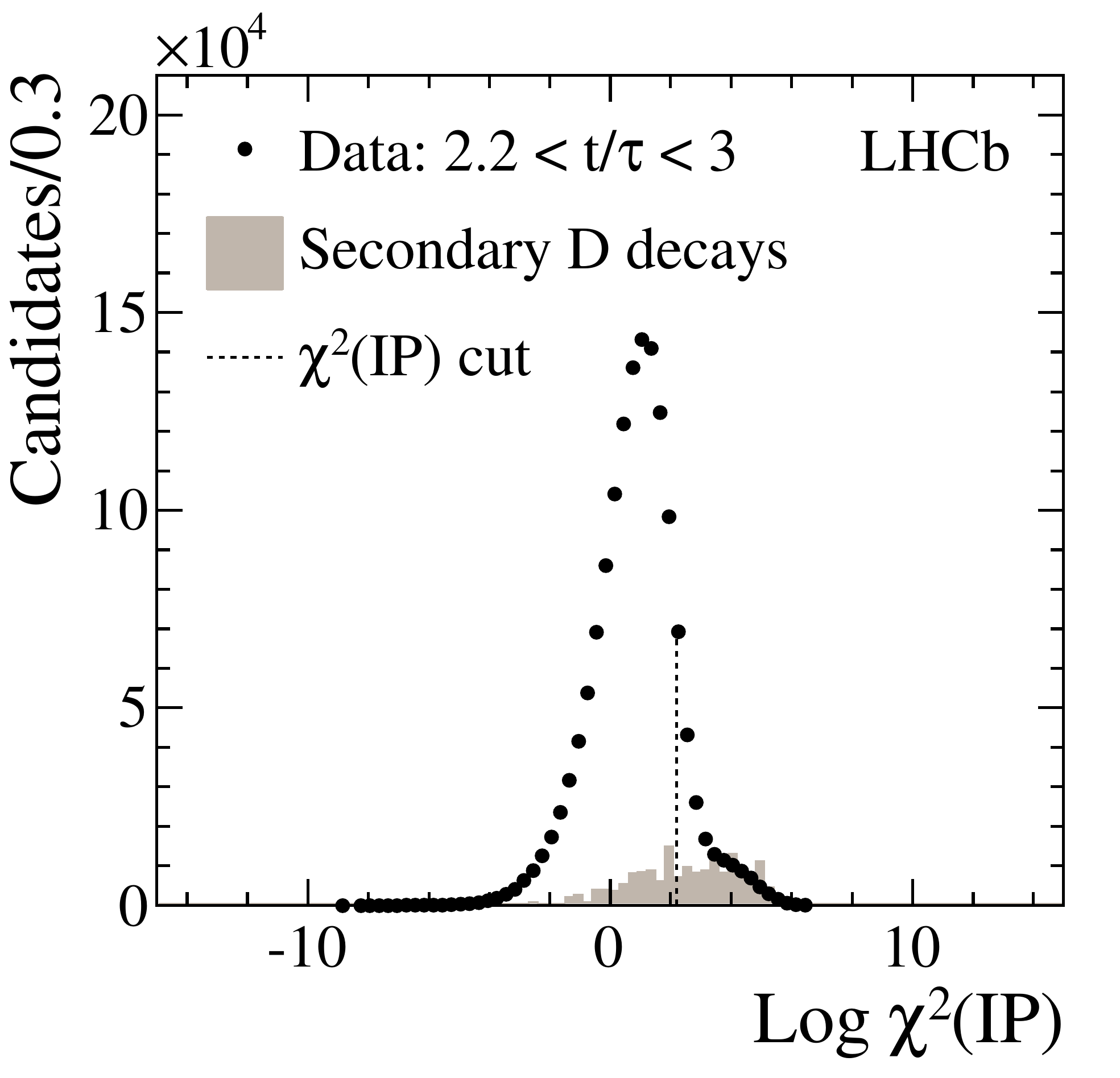}\hfil
\includegraphics[width=0.48\textwidth]{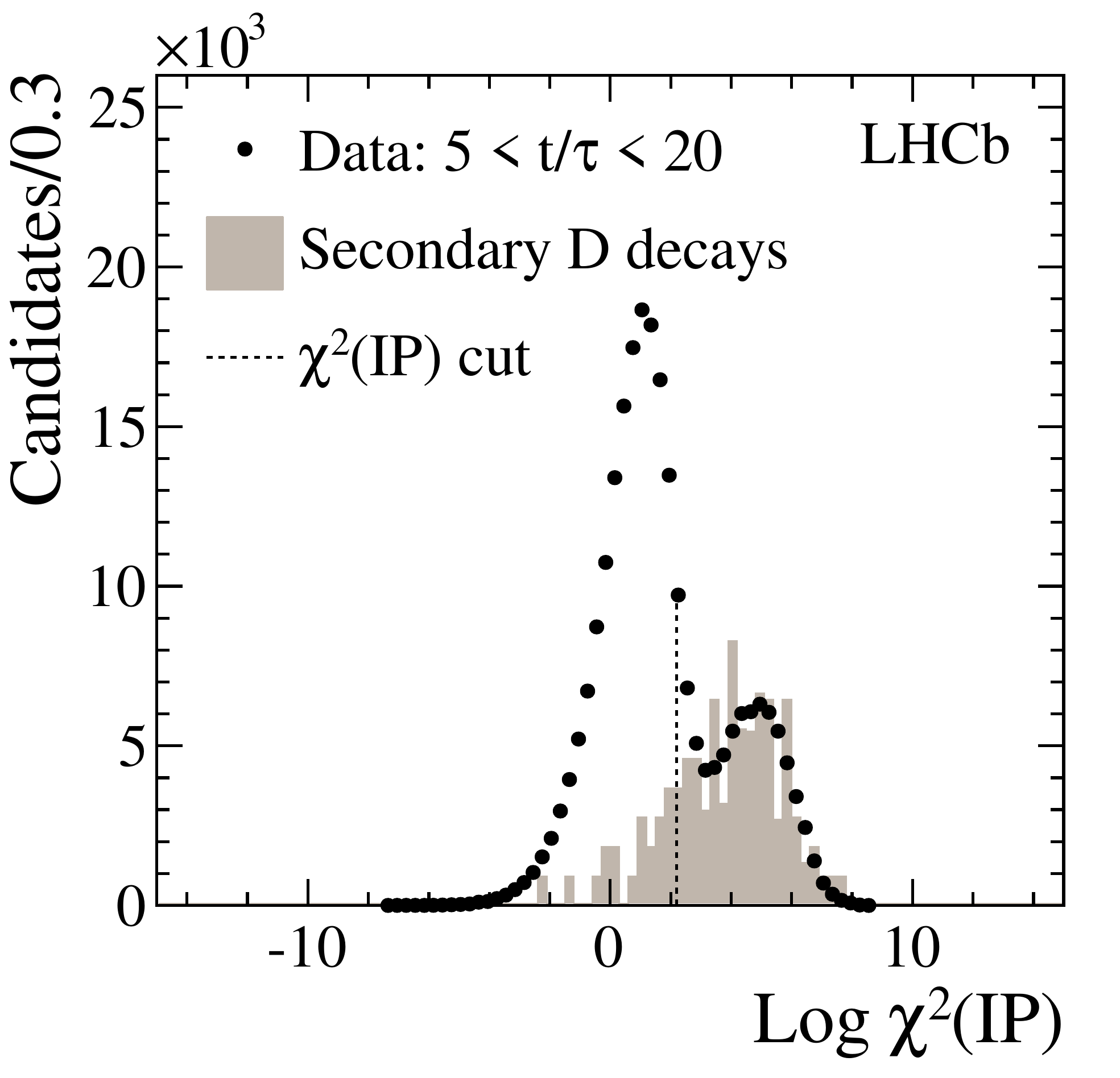}\\
\caption{Background-subtracted $\chi^2(\text{IP})$ distributions for RS \Dz candidates in two different decay-time bins. The dashed line indicates the cut used in the analysis; the solid histograms represent the estimated secondary components.}\label{fig:ipfit}
\end{figure*}

\begin{figure}[h]
\centering
\includegraphics[width=0.5\textwidth]{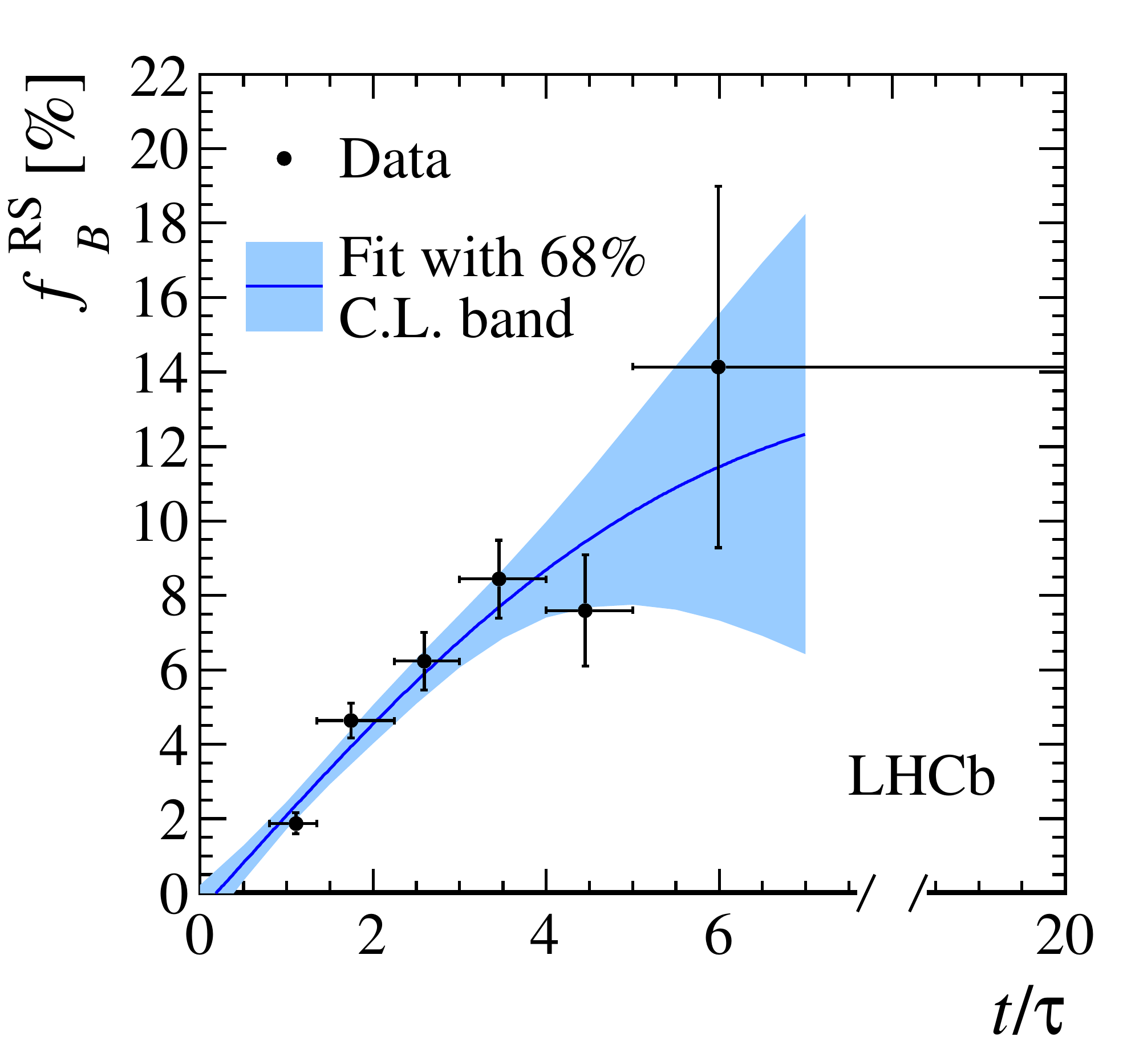}\\
\caption{Measured fraction of secondary decays entering the final RS sample as a function of decay time (points), with overlaid the projection of a fit to a sigmoid-like function (solid line).}\label{fig:secondary}
\end{figure}

A contamination of charm mesons produced in \bquark-hadron decays (secondary $D$ decays) could bias the time-dependent measurement, as the reconstructed decay time is calculated with respect to primary vertex, which, for these candidates, does not coincide with the \Dz production vertex. When the secondary component is not subtracted, the measured WS/RS ratio can be written as $R(t)\left[1 - \Delta_B(t)\right]$, where $R(t)$ is the ratio of promptly-produced candidates according to Eq.~\eqref{eq:true-ratio}, and $\Delta_B(t)$ is a time-dependent bias due to the secondary contamination. Since $R(t)$ is measured to be monotonically non-decreasing~\cite{hfag} and the reconstructed decay time for secondary decays overestimates the true decay time of the \Dz meson, it is possible to bound $\Delta_B(t)$, for all decay times, as
\begin{equation}\label{eq:Bbias}
0\leqslant \Delta_B(t)\leqslant f_B^{\rm RS}(t)\left[1-\frac{R_D}{R(t)}\right],
\end{equation}
where $f_B^{\rm RS}(t)$ is the fraction of secondary decays in the RS sample at decay time $t$. In this analysis most of the secondary \D decays are removed by requiring the $\chi^2(\text{IP})$ of the \Dz to be smaller than $9$.\footnote{The $\chi^2(\text{IP})$ is defined as the difference between the $\chi^2$ of the primary vertex reconstructed with and without the considered particle, and is a measure of consistency with the hypothesis that the particle originates from the primary vertex.} A residual $(2.7\pm0.2)\%$ contamination survives. To include the corresponding systematic uncertainty, we modify the fitting function for the mixing hypothesis assuming the largest possible bias from equation~\eqref{eq:Bbias}. The value of $f_B^{\rm RS}(t)$ is constrained, within uncertainties, to the measured value, obtained by fitting the $\chi^2(\text{IP})$ distribution of the RS \Dz candidates in bins of decay time (see figure~\ref{fig:ipfit}). In this fit, the promptly-produced component is described by a time-independent $\chi^2(\text{IP})$ shape, which is derived from data using the candidates with $t<0.8\tau$. The $\chi^2(\text{IP})$ shape of the secondary component (represented by the solid histograms in figure~\ref{fig:ipfit}), and its dependence on decay time, is also determined from data by studying the sub-sample of candidates that are reconstructed, in combination with other tracks in the events, as $B\to D^*(3)\pi$, $B\to D^*\mu X$ or $B\to D^0\mu X$. The measured value of $f_B^{\rm RS}(t)$ is shown in figure~\ref{fig:secondary}. Assuming the maximum bias could induce an over-correction which results in a shift in the estimated mixing parameters. We checked on pseudo-experiments, before fitting the data, and then also on data that such a shift is always much smaller than the corresponding increase in the uncertainty when the secondary bias is included in the fit.

\begin{figure}[h]
\centering
\includegraphics[width=0.5\textwidth]{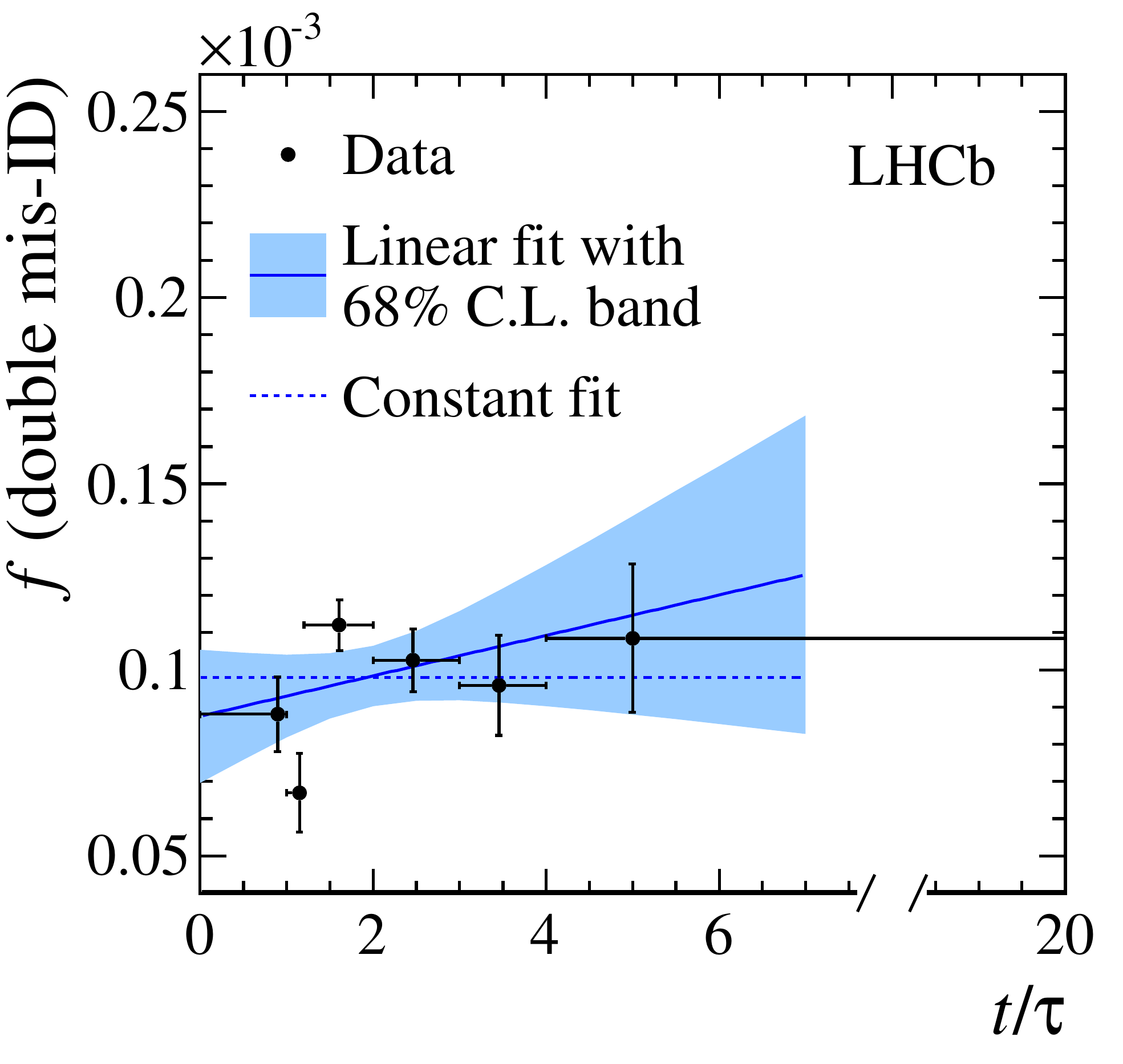}\\
\caption{Decay-time evolution of the number of doubly misidentified RS events observed in the \Dz mass sidebands of the WS sample normalized to the RS signal yield. The solid (dashed) line is the result of a fit assuming linear (constant) decay-time dependence.}\label{fig:peaking}
\end{figure}

Peaking background in \M, that is not accounted for in our mass fit, arises from \Dstarp decays for which the correct soft pion is found but the \Dz is partially reconstructed or misidentified. This background is suppressed by the use of tight particle identification and two-body mass requirements. From studies of the events in the \Dz mass sidebands, we find that the dominant source of peaking background leaking into our signal region is from RS events which are doubly misidentified as a WS candidate; they are estimated to constitute $(0.4\pm0.2)\%$ of the WS signal. From the same events, we also derive a bound on the possible time dependence of this background (see figure~\ref{fig:peaking}), which is included in the fit in a similar manner to the secondary background. Contamination from peaking background due to partially reconstructed \Dz decays is found to be much smaller than $0.1\%$ of the WS signal and neglected in the fit.

\subsection{Results}
\begin{figure}[ht]
\centering
\includegraphics[width=0.5\textwidth]{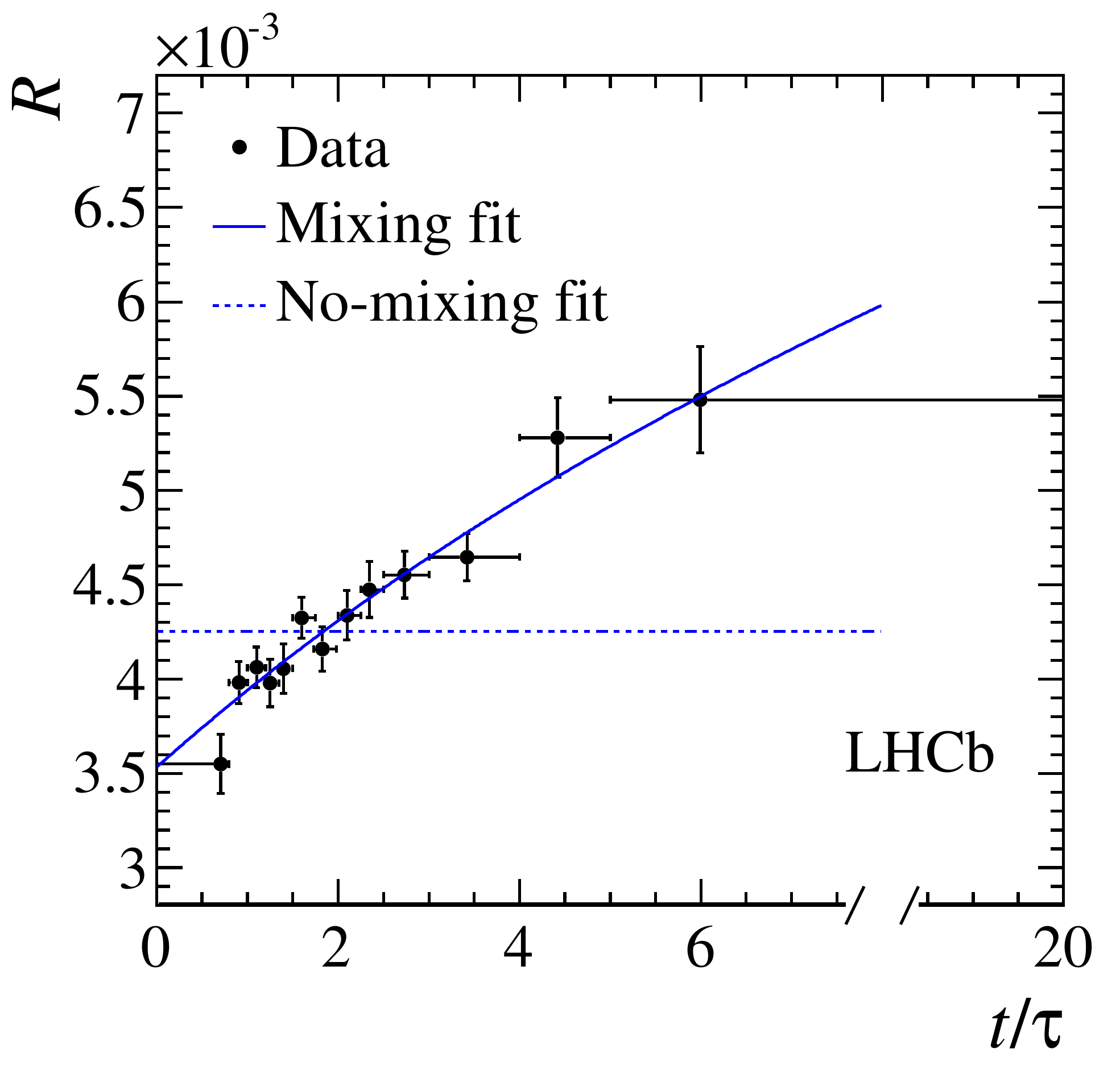}\\
\caption{Decay-time evolution of the ratio, $R$, of WS \mbox{$\Dz\to K^+\pi^-$} to RS $\Dz\to K^-\pi^+$ yields (points) with the projection of the mixing allowed (solid line) and no-mixing (dashed line) fits overlaid.\label{fig:results}}
\end{figure}

Figure~\ref{fig:results} shows the observed decay-time evolution of the WS to RS ratio, with the projection of the best fit result overlaid (solid line). The estimated values of the parameters $R_D$, $y'$ and $x'^2$ are listed in table~\ref{tab:results}.

\begin{figure}[ht]
\centering
\includegraphics[width=0.5\textwidth]{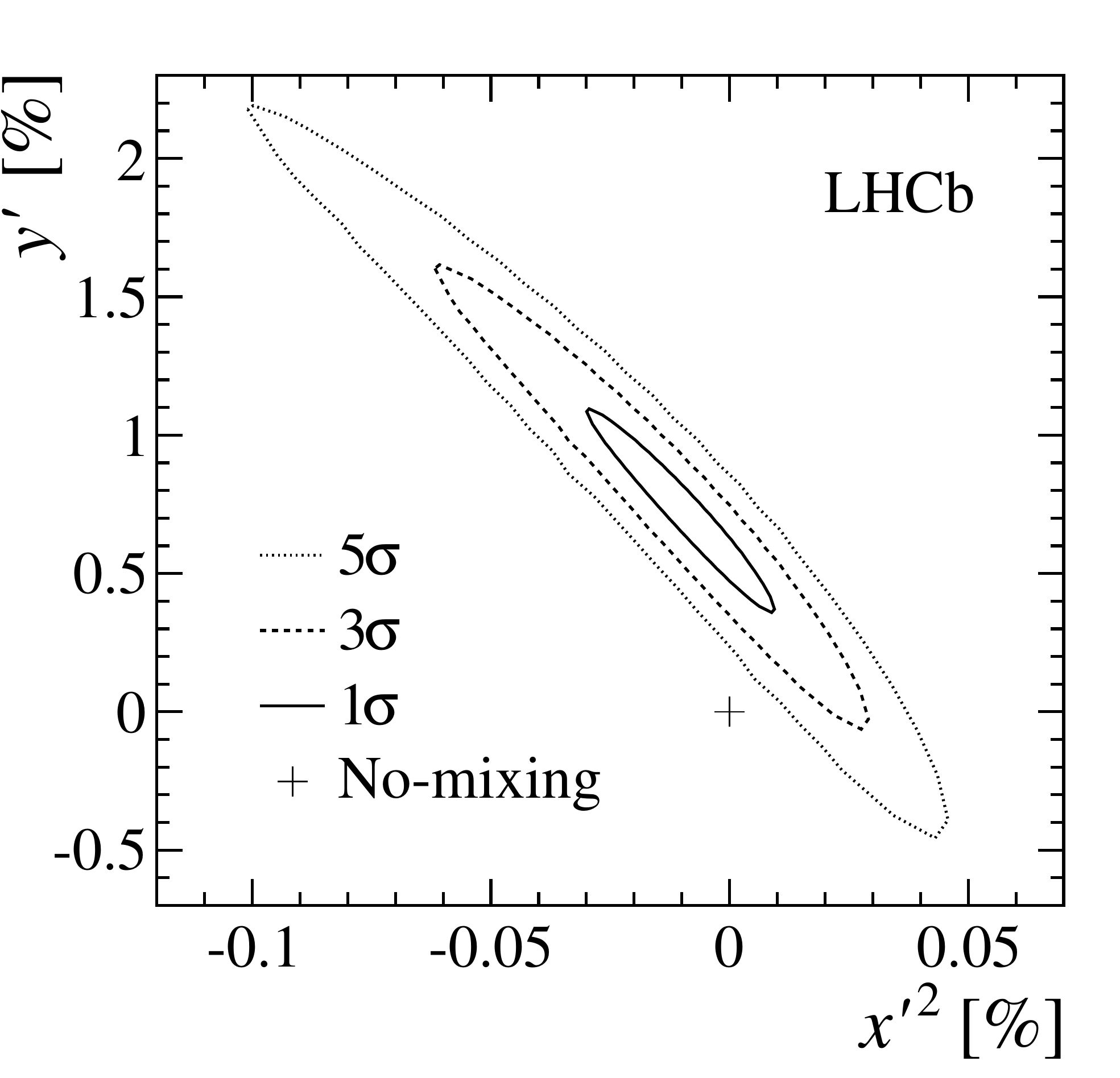}\\
\caption{Estimated confidence-level (CL) regions in the $(x'^2,y')$ plane for $1-\text{CL}=0.317$ ($1\sigma$), $2.7\times10^{-3}$ ($3\sigma$) and $5.73\times10^{-7}$ ($5\sigma$). Systematic uncertainties are included. The cross indicates the no-mixing point.}\label{fig:contours}
\end{figure}

\begin{table}[ht]
\centering
\caption{Results of the time-dependent fit to the data. The uncertainties include statistical and systematic sources; ndf indicates the number of degrees of freedom.\label{tab:results}}
\begin{tabular}{lcc}
\hline
Fit type ($\chi^2$/ndf) & Parameter &  Fit result ($10^{-3}$) \\
\hline
Mixing ($9.5/10$) & $R_D$ & $\quad3.52\pm0.15$  \\
 & $y'$ & $\quad7.2\pm2.4$ \\
 & $x'^2$ & $\,-0.09\pm0.13$\\
\hline
No mixing ($98.1/12$) & $R_D$ & $\quad4.25\pm0.04$\\
\hline
\end{tabular}
\end{table}

To evaluate the significance of this mixing result we determine the change in the fit $\chi^2$ when the data are described under the assumption of the no-mixing hypothesis (dashed line in figure~\ref{fig:results}). Under the assumption that the $\chi^2$ difference, $\Delta \chi^2$, follows a $\chi^2$ distribution for two degrees of freedom, $\Delta\chi^2=88.6$ corresponds to a $p$-value of $5.7\times 10^{-20}$, which excludes the no-mixing hypothesis at $9.1$ standard deviations. This is also illustrated in figure~\ref{fig:contours} where the $1\sigma$, $3\sigma$ and $5\sigma$ confidence regions for $x'^2$ and $y'$ are shown.

\section{Conclusions}
We measure the decay time dependence of the ratio between $\Dz\to K^+\pi^-$ and $\Dz\to K^-\pi^+$ decays using $1.0\invfb$ of data and exclude the no-mixing hypothesis at $9.1$ standard deviations. This is the first observation of $\Dz-\Dzb$ oscillations in a single measurement. The measured values of the mixing parameters are compatible with and have better precision than those from previous measurements \cite{Aubert:2007wf,Aaltonen:2007ac,Zhang:2006dp}, as shown in table~\ref{tab:comparison}.

\begin{table}[h]
\centering
\caption{Comparison of our result with recent measurements from other experiments. The uncertainties include statistical and systematic components.\label{tab:comparison}}
\begin{tabular}{lccc}
\hline
Exp. & $R_D$ ($10^{-3}$)  & $y'$ ($10^{-3}$)  & $x'^2$ ($10^{-3}$)  \\
\hline
LHCb 2011 & $3.52\pm0.15$ & $7.2\pm2.4$ & $-0.09\pm0.13$ \\
BaBar \cite{Aubert:2007wf} & $3.03\pm0.19$ & $9.7\pm5.4$ & $-0.22\pm0.37$ \\
Belle \cite{Zhang:2006dp} & $3.64\pm0.17$ & $0.6^{+4.0}_{-3.9}$ & $0.18^{+0.21}_{-0.23}$ \\
CDF \cite{Aaltonen:2007ac} & $3.04\pm0.55$ & $8.5\pm7.6$ & $-0.12\pm0.35$ \\
\hline
\end{tabular}
\end{table}

The result is still statistically limited and can be improved thanks to the already available $2.1\,\text{fb}^{-1}$ sample of $pp$ collisions recorded by \lhcb at $\sqrt{s}=8\,\tev$ during 2012. This additional sample will also allow to search for \CP violation in charm mixing with sensitivities never achieved before.

\end{document}